# Bus Factor In Practice

Elgun Jabrayilzade[*]
elgun@bilkent.edu.tr
Bilkent University
Ankara, Turkey

Mikhail Evtikhiev[*]
mikhail.evtikhiev@jetbrains.com
JetBrains Research
Saint Petersburg, Russia

Eray Tüzün
eraytuzun@cs.bilkent.edu.tr
Bilkent University
Ankara, Turkey

Vladimir Kovalenko
Vladimir.Kovalenko@jetbrains.com
JetBrains Research
Amsterdam, The Netherlands

## ABSTRACT
Bus factor is a metric that identifies how resilient is the project to the sudden engineer turnover. It states the minimal number of engineers that have to be hit by a bus for a project to be stalled. Even though the metric is often discussed in the community, few studies consider its general relevance. Moreover, the existing tools for bus factor estimation focus solely on the data from version control systems, even though there exists other channels for knowledge generation and distribution. With a survey of 269 engineers, we find that the bus factor is perceived as an important problem in collective development, and determine the highest impact channels of knowledge generation and distribution in software development teams. We also propose a multimodal bus factor estimation algorithm that uses data on code reviews and meetings together with the VCS data. We test the algorithm on 13 projects developed at JetBrains and compared its results to the results of the state-of-the-art tool by Avelino et al. against the ground truth collected in a survey of the engineers working on these projects. Our algorithm is slightly better in terms of both predicting the bus factor as well as key developers compared to the results of Avelino et al. Finally, we use the interviews and the surveys to derive a set of best practices to address the bus factor issue and proposals for the possible bus factor assessment tool.

## 1 INTRODUCTION

Software projects are rarely developed by a single person. According to the ISBSG repository [1], the average size of a software development team, averaged over time, is 7.9 members, and the median team size is 5 [20]. In collective work, it may be nontrivial to track the knowledge distribution in the team. Tracking knowledge distribution is important, as e.g. work by the engineers with low expertise on a given artifact is known to be more bug-prone [6].

Knowledge tracking may be further impeded by the changes in team membership. Departure of the key project members can lead to a situation when a significant part of the project is poorly understood by the remaining project members. This staff turnover can result in project stalling. For example, Avelino et al. [4] have found that out of 1,932 open source projects 16% of the projects have faced the departure of all key engineers, and in only 41% of these projects, the development has been continued by other engineers. Learning how the knowledge about the project is distributed (and acting on that knowledge) can help to identify projects with high existential risks. This information enables a team or its manager to manage risks related to sudden departure of engineers.

One metric that tracks project stalling risk is the *bus factor*. The bus factor is the minimal number of engineers that have to leave for a project to stall. We call the stalling "bus factor problem". The bus factor number is usually supplemented by a list of the key engineers who are responsible for it [5, 10, 18].

There are two possible issues with using bus factor as a health metric for a project. First, even though a significant share of the open-source projects have experienced bus factor problem, it is unclear whether this result was undesirable or it was simply end of life for a project no longer needed. Engineers may also perceive bus factor problem risk as unimportant or improbable as compared to other problems that arise in collective development. While such a perception does not affect the significance of the metric, it may mean that the community is more interested in other project health metrics. It may also mean that for a given project the engineers may not address the possible risks of the bus factor problem, as other more urgent or important problems should be addressed first.

Second, while the bus factor metric is easy to comprehend, it is unclear how to compute it without asking the stakeholders. Asking stakeholders directly is a feasible approach for small projects (and can help in tuning the bus factor computation algorithms). For large and distributed projects there may be no stakeholder with a full picture of what is going on, so a bus factor computation algorithm can provide information useful for risk mitigation.

A standard approach to evaluate the bus factor hinges on collecting information about the project from the available sources. This data is used to estimate the bus factor. In the existing papers, the bus factor is estimated solely from the data collected from the version control systems (VCS). However, the knowledge is shared and created not only by writing code [13]. It is possible to contribute to a project or to share knowledge about it through code reviews, discussions at meetings, chats, mailing lists or issue trackers. Leaving out these knowledge distribution modes may result in an inaccurate bus factor estimation. A particular example is a team led by a senior engineer, who hardly writes code, but participates in discussions and performs code reviews. In this case, VCS-based bus factor algorithm can incorrectly label this senior engineer not to be a key member of a project, while in practice they may have the most knowledge about the project. A bus factor algorithm that in addition to VCS uses project members' communications data, and code reviews data, could give better estimation of knowledge distribution.

---

[*]Elgun Jabrayilzade and Mikhail Evtikhiev contributed equally.

These observations have motivated us to pose the following research questions:

**RQ0** Do engineers believe that the bus factor is an important issue in collective development?
**RQ1** How the bus factor metric can be used in practice?
**RQ2** What features a bus factor assessment algorithm should have?
**RQ3** How accurate are the results on the bus factor and key project members, as computed by our algorithm and as estimated by the baseline algorithm?

To answer these research questions, we did a two-prong research. To answer RQ0 and RQ1, we have first conducted a set of exploratory interviews to validate general concepts about bus factor perception in the community. Using the results of the interviews, we have then created a general bus factor survey of 269 engineers that was in part designed to collect opinions on the bus factor concept importance and relevance. The survey was available for both JetBrains employees and engineers that don't work at JetBrains.

To answer RQ2 and RQ3, we have performed the following steps:

- To validate our choice of the bus factor algorithm, in the bus factor survey we collect opinions on a set of questions about the algorithm (full list can be found in Section 4).
- Together with the bus factor survey, we performed a case study on the projects developed at JetBrains. The engineers were asked to provide their estimates on the bus factor and the list of key engineers for the projects they work on. In total, we have collected estimates for 13 projects hosted at the internal JetBrains' Space instance. JetBrains Space is an all-in-one solution for project management and software development for software projects and team, that has integration with Google Calendar, VCS, code reviews, and has API that allows retrieving information about users' activities.
- We have then estimated the bus factor of these projects using our algorithm and the algorithm of Avelino et al. [5], which we have used as a baseline algorithm. We validated the results of our algorithm and the algorithm of Avelino et al. against the estimates collected in the survey.

To use the tool, we gather the data on code reviews, meetings, and commits from the JetBrains Space instance used by JetBrains.

Our contributions in this paper are the following. First, we propose a multimodal bus factor algorithm to estimate the existential risks of a project from a diverse set of data that can be harvested from JetBrains Space or other platforms that host multiple types of data. Second, we carry out a survey of engineers that validates the results of our algorithm and compares it to the baseline algorithm of Avelino et al. [5]. Third, we carry out a survey and a set of exploratory interviews that show the importance of the bus factor through the lens of the project members' experience. Using the survey and the interviews, we derive a set of recommendations and requests for the bus factor tool and check our assumptions about the composition of the bus factor algorithm.

## 2 BACKGROUND
### 2.1 Bus factor as a collective development problem

Most of the software projects are developed by a group of people. Knowledge about the project can be derived from the code, but reading, improving, and supporting new code is a very hard task for an engineer. The files abandoned by their original developers remain abandoned for a long time [15]. A departure of a significant number of key engineers can result in an abandonment of an important part of the codebase. This departure may lead to project stalling.

Bus factor (also known as truck factor, bus number, lottery factor, etc.) was defined by Coplien as the minimal number of the developers that would have to be hit by a bus before the project is stalled[1] [9]. Smaller values of the bus factor correspond to higher existential risks for projects. Higher values (as compared to the team size) correspond to a relatively even distribution of knowledge, so a departure of a project member should have a lesser impact.

Bus factor problem is a collective development problem that happens when the project bus factor becomes zero. The notion of collective development problem is close to the notion of the community smell suggested by Tamburri et al. [22]. A community smell is a set of sub-optimal organizational structures that lead to the emergence of both social and technical debt [11]. The difference between the two concepts is that community smells are not "show-stoppers" [23], but rather reflect circumstances that with time manifest in additional project cost. A collective development problem is a problem perceived as a factor that already hampers project development. It is possible that every collective development problem is a result of aggravation of one or several community smells, and for some of the problems discussed in this paper we find plausible relations or one-to-one correspondence. Proving this assertion is beyond the scope of this paper.

There have been many advances in the study of community smells in the past few years. A set of community smells have been first identified in a paper by Tamburri et al. [21]. Tamburri et al. [23] have created an automated tool that detects the so-called Organizational Silo, Black Cloud, Lone Wolf, and Radio Silence community smells. Palomba et al. [17] have studied the relationship between community smells and code smells to propose a code smell intensity prediction model that relies on both technical and community-related aspects. Palomba et al. [16] and Almarimi et al. [3] have suggested machine learning models that predict the existence of the community smells from the social and organizational patterns of the developers' community. Almarimi et al. identify the Truck Factor smell related to the bus factor notion and learn to predict it. Catolino et al. [7] have studied the relationship between various socio-technical patterns and community smells from the statistical point of view. They found that communicability is important in managing community smells, while broadening the collaboration network does not always help. Finally, Catolino et al. [8] did an empirical study on the relevance of community smells in practice and the refactoring approaches that can help to eliminate the smells.

---
[1]Different definitions of the bus factor exist in the literature; sometimes words "incapacitated" or "abandoned" are used instead of "stalled".

## 2.2 Bus factor algorithms

It is nontrivial to translate the bus factor definition into a bus factor estimation algorithm. As it may be hard for the project members to compute the bus factor for a large and distributed project they work on, an algorithm estimating the bus factor of a project from the data about the project could decrease the project stalling risks.

There exist several algorithms for bus factor estimation, which we describe below. In each of these algorithms, the bus factor is estimated by studying the distribution of the knowledge about the project derived from the VCS logs and data.

Zazworka et al. [25] suggested a highly configurable algorithm first implemented by Ricca et al. [18]. The algorithm starts with a list of files in a project, a list of developers who are considered to be knowledgeable for each of the files (both are mined from the VCS history), and a threshold value $X$. The algorithm finds the minimal set of the developers such that these developers belong to more than X% of the project files. This developer set is then identified as a key engineer set and the bus factor is the size of the set. A standard setup of Ricca et al. [18] was to consider every developer who has edited the file to be knowledgeable about the file, and the threshold was varied from 50% to 70%. The authors have applied the algorithm with various thresholds, finding that:

- The threshold value affects the computed bus factor.
- The idea that every person who did at least one commit to a file has knowledge about a file may be too strong.
- There are "update" commits that cover many files but probably do not signify that the committer has knowledge about them.
- The algorithm performs well for the small projects, but seems to be problematic for projects with > 30 committers.

Cosentino et al. [10] suggest a set of algorithms to compute the bus factor. The authors define the notion of primary and secondary developers. The file is abandoned when none of the primary or secondary developers who worked on it are present in the project. The bus factor problem happens when a certain amount of files is abandoned. The analysis can be carried out at either file-level or line-level of granularity. The authors suggest four different metrics (M1 to M4) for measuring contribution. The contributions are tracked at the commit level, and the ownership is tracked per file.

- M1 assigns all knowledge to the last contributor.
- M2 considers every change with equal weight (e.g., if A committed to a file twice and B committed once, A has 2/3 knowledge and B has 1/3).
- M3 is the same as M2, but consecutive changes are considered to be one change (in A -> A -> B commit history, both A and B have knowledge 1/2).
- M4 adds weights to the changes; for a file with N commits, the oldest commit has weight 1, the second-oldest has weight 2, and so on.

Applying any of these metrics for a file yields a list of contributors with their shares of contribution to the file, with the shares summing up to 100%. For a file that has been edited by $N$ contributors, primary developers are those who have done at least $100/N$% of edits and secondary developers have done $100/N\% > x > 50/N\%$ of edits.

Rigby et al. [19] suggest an algorithm for computing knowledge loss, that can be adapted to estimate the project bus factor. Knowledge loss happens when the developer who owns a line of code leaves the project. It is calculated as the number of the abandoned files. A file is abandoned when > 90% of its lines are owned by the engineers who have left the project. The authors suggest computing historical knowledge loss distribution, accounting for the pieces of abandoned code that were easily picked up by other developers. Rigby et al. [19] use the historical data to estimate the percentage of knowledge at risk (knowledge that can be lost with > 5% probability) and risk from unexpected high losses. This approach quantifies risks explicitly and thus may be useful for risk mitigation and the risk/cost/benefit analysis. They also suggest finding successors for the abandoned code to mitigate the existential risks.

Fritz et al. [13] have suggested a Degree of Knowledge metric to track the code ownership distribution in the files. Degree of Knowledge (DOK) is a composite metric given by a linear combination of Degree of Authorship (DOA) and Degree of Interest (DOI) metrics. Degree of Interest metric was first suggested by Kersten et al. [14] and is computed based on the amount of engineer's interactions with the element. The impact of an interaction decays with time, so more recent interactions are given higher weights. Degree of Authorship is a metric computed from the number of commits made to a file, and also takes into account who created the file:

$$DOA(e, f) = 3.293 + 1.098 FA + \\ + 0.164 DL - 0.321 \log(1 + AC), \quad (1)$$

where $FA$ (first authorship) is 1 for $f$ file creator and 0 otherwise, $DL$ is the number of commits to the file $f$ made by the engineer $e$, and $AC$ is the number of commits to the file $f$ made by the other project members. Only the contributions made in the last 90 days are considered in the DOA. Their analysis shows that the DOI is not correlated to any of the variables in the DOA metric, suggesting that DOI has a predictive role. The authors do not estimate the bus factor. However, the finding of Fritz et al. that DOI is a measure of a developer's knowledge uncorrelated to the commit-based DOA metric indicates that considering VCS history may not be enough to capture the knowledge distribution, and the bus factor of a project.

Avelino et al. [5] have suggested a bus factor computation algorithm based on the Degree of Authorship (DOA) metric suggested by Fritz et al. [13]. Their algorithm first computes the DOA of each of the engineers for every file in the project according to the (1). An engineer $e$ is an author of a file $f$ if $DOA(e, f) > 3.293$ and

$$DOA(e, f) > 0.75 * \max_e DOA(e, f), \quad (2)$$

where 3.293 is the constant equal to the free term of (1), and $\max_e DOA(e, f)$ is the highest DOA for file $f$ for all project members. A file is considered abandoned if all its authors have left the project. The algorithm of Avelino et al. [5] takes the list of files with the DOAs of the engineers who have worked on them, and iteratively moves the top author[2] from the list of present developers to the key engineers list. The bus factor problem is considered to happen when more than 50% of files have been abandoned, and the algorithm yields the bus factor together with the key engineers list.

---

[2] An author is a top author if they author more files than anyone else

Ferreira et al. [12] have carried out a comparative study of various bus factor algorithms. In their study, they consider the algorithms of Rigby et al. [19], Zazworka et al. [25], Avelino et al. [5] and Cosentino et al [10]. In addition, they consider two algorithms of Yamashita et al. [24] that identify the core developers (the core developer definition is close to the key engineer definition). Ferreira et al. [12] validated the algorithm results on a dataset of 35 open-source projects. The authors of the study presented the developers of the open-source projects with the data produced by the algorithm of Avelino et al. [5] and asked whether they agree with the algorithm assessment. Authors have shown that the algorithms of Avelino et al. [5] and Cosentino et al. [10] are the most accurate algorithms for both bus factor and key engineer estimation, with the Avelino et al. algorithm performing slightly better. However, all studied algorithms perform worse on projects with high bus factor. The authors have also shown that the algorithm of Avelino et al. [5] is better at determining the bus factor than the core developers algorithms. Finally, the authors have considered the projects where key engineers (as identified by the project members) have done very few commits. They have reached out to the developers of these projects to ask how did the bus factor developers with very few commits have contributed to the projects. The respondents have reported that social interactions, code reviews, test writing, documentation writing, and tool support have been important ways for them to contribute to the projects.

Almarimi et al. [2, 3] have created a csDetector tool that detects several community smells, and the Truck Factor smell in particular. The smell is determined by a machine learning model on a yes/no basis and no additional information is presented. In [3], authors have compared their tool to the tool of Avelino et al. [5]. The algorithm of Avelino et al. [5] has shown 0.84 accuracy on the dataset supplemented by authors, while csDetector tool has 0.97 accuracy. Almarimi et al. [2] define that the algorithm of Avelino et al. finds a Truck Factor smell, if the departure of two or fewer contributors results in the abandonment of more than 40% of the files. As csDetector analyzes a significantly different feature of a project, we did not compare our tool with theirs.

Existing bus factor estimation algorithms rely solely on the VCS data to estimate the bus factors. This approach allows collecting a plethora of data from open-source projects. However, the study of Ferreira et al. [12] shows that for a third of the projects from the dataset some of the real key engineers committed to the project not by writing code, but by doing code-related activities that are not captured in the repository. Moreover, Fritz et al. [13] show that considering just the information from the repository is not enough to capture all the data about the code ownership distribution. The information about the developers' interactions with the code that are not logged in the VCS logs is relevant and uncorrelated with VCS data. All these findings have motivated us to try creating a multimodal bus factor algorithm that incorporates code reviews, VCS data, and meetings data harvested from JetBrains Space.

## 3 BUS FACTOR ALGORITHM AND TOOL
### 3.1 Bus Factor Algorithm

We took the DOA formula designed by Fritz et al. [13] as a reference and adjusted it to incorporate contribution decay, code reviews, and meetings. The new formula is given below:

$$DOA_i = 3 \cdot FA + \sum_j DL_i^j + \frac{1}{2} \sum_j RV_i^j + \sum_l min(1, \sum_j \frac{MT_i^{l,j}}{MTE})) +$$
$$+ 2.4 \log(1 + \sum_k \sum_j DL_k^j) + 1.2 \log(1 + \sum_k \sum_j RV_k^j) -$$
$$- 2.4 \log(1 + \sum_{k \neq i} \sum_j DL_k^j) - 1.2 \log(1 + \sum_{k \neq i} \sum_j RV_k^j)$$
$$FA = \exp(\frac{-t_{FA}}{S}), DL_i^j = \exp(\frac{-t_{DL^j}}{S})$$
$$RV_i^j = \exp(\frac{-t_{RV^j}}{S}), MT_i^j = \exp(\frac{-t_{MT^j}}{S})$$

The $DOA_i$ represents the contribution of an engineer to a specific file. $FA$ represents the first authorship and can be 1 or 0. $DL_i^j$ is the number of commits made by an engineer $i$ to the file $j$. Similarly, $RV_i^j$ is the number of reviews done by an engineer $i$ to the file $j$. $MT_i^{l,j}$, on the other hand, represents the number of minutes that the engineer $i$ spends on $j^{th}$ meeting about commit $l$. $MTE$ is the maximum effective time in a meeting we set at 240 minutes. In other words, an engineer can get a maximum of 240 minutes from meetings on an arbitrary commit. The knowledge contributions from all parts decay exponentially according to the number of days passed since the contribution made. We set inverse decay speed $S = 220$, which means that the knowledge from a contribution halves in about five months. An engineer is said to be a major contributor of a file if the $DOA >= 1$ and the normalized $DOA >= 0.75$.

The bus factor calculation is based on the algorithm developed by Avelino et al. [5]. The top authors are removed iteratively until the current engineers' knowledge covers less than half of the files. The number of removed engineers represents the bus factor.

### 3.2 Bus Factor Calculation Tool

*3.2.1 Data retrieval.* JetBrains uses the Space tool that is an all-in-one solution for managing the software development lifecycle. The tool has HTTP API endpoints for retrieving data. We retrieve the list of the employees (names, emails, profile links), projects, Git histories of the project repositories, code reviews (status, reviewers, commits, date), and meetings (participants, duration, date).

*3.2.2 Data cleaning.* We extract committers' emails and names from the Git histories and merge accounts having the same email since one can use multiple Git accounts. We also map committers to their Space profiles to track their knowledge contributions from code reviews and meetings. Committers without Space profiles get knowledge contribution only from commits. We filter code reviews to include only the pull requests merged to the codebase. To include only relevant meetings that can contribute to the knowledge, we exclude the meetings containing *seminar*, *reading*, or *random* keywords in their descriptions.

3.2.3 *Data processing.* We use the JGit[3] library to parse Git histories of the given repositories. The bus factor is calculated for each branch separately. The Git graphs of the branches are traversed via depth-first search. Contributions from each commit are calculated by extracting the difference between the current commit and its parent commit. Contributions from merge conflicts are handled by taking the intersection of differences between the commit and its two parents. The file renames are tracked, and such commits do not contribute toward the knowledge on file. Next, the code reviews are processed, and the contributions are counted for each reviewer-commit pair unless the reviewer is the same as the committer (self-review). Finally, to include the contributions from meetings as well, we extract the meetings that might be related to a given commit. We assume that the related meetings are the ones which were attended by the committer and occurred within one week of the commit. Then, the number of minutes that the participants spend on those meetings is counted as a contribution towards the knowledge.

## 4 STUDY DESIGN

In addition to the tool development, we have conducted three human studies. Study 1 is a set of exploratory interviews that investigate the communication and coordination difficulties engineers may face in their jobs. Study 2 is a survey that targets engineers' opinions about the factors that influence the project bus factor, and their opinions about bus factor importance and using the bus factor to evaluate project health. Finally, Study 3 is an additional part of a survey only available to the JetBrains engineers which collected their opinion on the bus factor of the projects they work on.

### 4.1 Exploratory interviews

Our exploratory interviews pursued two major goals. First, we wanted to get a broader picture of what collective development problems software engineers encounter in practice. This was done in order to create survey questions to study the relative importance of the bus factor as compared to the other common collective development problems. We identified group of collective development problems different from the community smells reported in [17], for the common ones we use the original names suggested by Palomba et al. The difference may be explained by the difference in the scope of the research conducted by Palomba et al. [17] and by us. While Palomba et al. [17] have considered the effect the community smells exhibit on a particular code smell, we have focused on the collective development problems that impact the whole project.

Our second goal was to collect a set of cases on the collective development problems and bus factor problem cases in particular. The responses allowed us to get a deeper understanding of why and how these problems occur, and how the engineering teams resolve them. We also learned why these problems were not prevented and what respondents do now to avoid them.

*Recruitment.* For our interviews, we have recruited 12 engineers working at JetBrains and other companies who have at least 3 years experience of working in the IT teams. Participation in the study was voluntary and the participants received no compensation.

*Participants.* We have recruited participants with various working experiences: engineers, team leads, startup founders, project managers. The participants also reported working in a variety of roles, including developers, data scientist, QA, SRE, support engineers. The participants have 3 to 18 years of work experience (median: 12.5). Participants worked in various kinds of companies including product companies, outsourcing companies, and startups.

*Interview study protocol.* We have conducted semi-structured interviews remotely, and have recorded these interviews. First we have asked the participants about their work experience as an IT professional in general. We then asked them about coordination and organization problems that they have witnessed or were wary of, focusing on the problems that are **not** due to the personal differences but can rather be attributed to some ineffective team processes. We have followed up with a deeper discussion of a recent particular problem the respondent has witnessed. We focused on the origins of a problem and the measures respondent and their team has taken to resolve the problem and prevent it from repeating. If a respondent has reported facing several problems recently, we focused on the problem that was the closest to the bus factor problem. If a respondent did not report any problem that was close to the bus factor scenario, we have asked them directly whether they have witnessed a bus factor problem at any time during their work experience. Finally, we have asked the respondents if some kind of tool could help them to mitigate the bus factor risks and prevent the bus factor problem. We also discussed possible features of this tool with the participants. The interviews were from 60 to 90 minutes long, and the sessions were conducted with a single respondent and either one or two interviewers.

*Informed consent.* The participants received an explanation of the general interview process before the interview and gave explicit consent for participating in the interview.

*Analysis.* Mikhail Evtikhiev has processed the interviews and the transcripts. First, we have processed general demographic information, which is presented in Table 1. We have assigned short descriptions to categorize reported general problems and various development shortcomings that may lead to a bus factor problem. The collaborators have participated in several meetings to refine the discovered categories and discuss them.

*4.1.1 Results.* The data on the participants' demographics is presented in Table 1.

**Bus factor relative importance** Using the interview results, we selected the following collective development problems to ask about in the survey:

> **Bus factor.** In the survey, we describe the bus factor as "Nobody understands the code base of a crucial part of a project". Many respondents have reported they had to revive a project with BF = 0 (IP2, IP4, IP5, IP6, IP7, IP10, IP11, IP12). The bus factor problem emerges when the project has BF = 1 due to some crucial subsystem having exactly one engineer working on it. The bus factor problem then usually happens when this key engineer suddenly leaves the project, and the rest of the team has to pick up the development. This engineer is usually very knowledgeable and may develop the subsystem in a peculiar way, so it is hard for their colleagues to continue the development. Restoring ownership of the

---
[3]https://www.eclipse.org/jgit/

| ID | Current Roles | Experience (years) | Prior Roles |
|---|---|---|---|
| IP1 | Software Developer | 6 | Team Lead, CTO |
| IP2 | Software Developer | 3 | Not Applicable |
| IP3 | SRE, QA | 15 | Team Lead, Project Manager |
| IP4 | Support Engineer, QA | 10 | Software Developer |
| IP5 | Team Lead | 13 | Software Developer |
| IP6 | Team Lead | 18 | Software Developer |
| IP7 | Team Lead | 6 | Software Developer |
| IP8 | Software Developer, Project Manager | 18 | Team Lead, Tech Lead |
| IP9 | QA | 16 | Engineer |
| IP10 | Team Lead | 10 | Software Developer |
| IP11 | Data Scientist | 16 | Software Developer, Team Lead |
| IP12 | Team Lead | 12 | Data Scientist |

Table 1: Participants of the exploratory interviews.

QA = Quality Assurance Engineer. SRE = Site Reliability Engineer. CTO = Chief Technology Officer.

project is hard since the departed engineer usually provides answers with a high level of abstraction, while low-level abstraction is required (IP7). In some cases it took up to 6 person-months to get the project back on track (IP10), and certain subsystems even had to be rewritten from scratch (IP5, IP7).

**Code red.** In the survey, we describe the code red as "A crucial part of a project depends on too few people and the project stalls when they are sick or go on vacation". This definition is close to the definition of [17] and the problem is related to the bus factor. Most of our respondents have reported they worked on a project with code red problem (IP1, IP2, IP4, IP5, IP6, IP8, IP9, IP10, IP12). While, according to our respondents, code red may not always be an acute problem, some of them explicitly mentioned situations when corresponding key developers were not able to help and the productivity was reduced (IP2, IP5, IP8, IP10, IP12). Moreover, this situation presents risks of the bus factor problem.

**Vague responsibility.** In the survey, we describe vague responsibility as "Lack of concrete responsibility, so that it is unclear who is responsible for a particular feature or piece of code base".

**Dissensus.** In the survey, we describe the dissensus as "It is hard for the team to agree, what approach to choose to solve some technical problem". This problem is very close to the *Dissensus* community smell studied by Palomba et al. [17].

**Secret problems.** In the survey, we describe the secret problems as "A member of a team has significant technical problems but doesn't disclose problems to their colleagues, so when the release approaches, the team suddenly finds out that some feature is not ready".

**Lack of documentation.** In the survey, we describe the lack of documentation as "Lack of documentation for a project, which makes it hard for newcomers to start working on the project or figure out the details".

**Broken fix.** In the survey, we describe the broken fix as "An engineer tries to fix somebody else's code for their needs, but breaks some functionality used by other team members". This problem is similar to the *Lone Wolf* and *Dispersion* community smells studied by Palomba et al.

Respondents have suggested several reasons why the collective development problems are not addressed or fixed, which include:

- Inability to reach consensus within the team on how to resolve the problem or whether it is a problem at all (IP10).
- Lack of people to resolve issues faced by the team (IP12). This results in team having to triage the issues and collective development problems tend to grow if left unattended.
- Having people from outside the team in the decision-making loop (IP2, IP6).
- Overshadowing of the collective development problems by the technical problems perceived to be more urgent (IP11).

### 4.2 General bus factor survey

We have conducted a general bus factor survey that has pursued two goals. First, we wanted to validate our general concepts about knowledge distribution and decay in a project and how it impacts the bus factor of a project or its parts. We also wanted to validate the credibility of the observations we made during the exploratory interviews and to quantify the importance and occurrence of various problems that may lead to the bus factor scenario.

*Survey protocol.* Our survey can be broadly split into several parts. The first part includes several demographic questions about the number of years respondents work in IT, the number of projects they participated in as a line worker and as a leader of some group of people, and whether they know what bus factor is and have ever worked on a project where bus factor was tracked. We have filtered the survey results to include engineers with the current job title being Developer, Data scientist, Team Lead, Tester, DevOps, PM, CTO or related to one of these groups.

The second part was related to the perception of the collective development problems and their relative importance and frequency as compared to the bus factor scenario. In it, we have asked the engineers, how often do they encounter some of the collective development problems (see 4.1.1), and how significant the problems are. The order of problems presented was shuffled. For all the questions in the second part we have used a 5-point Likert-type scale. We have presented 7 activities in the survey.

The third part involved questions on how the knowledge is distributed, generated, and forgotten. In particular, we have asked the participants to rank the relative importance of different modes of knowledge generation and distribution, see Table 2. We have also asked the participants to estimate the maximal percentage of knowledge-related files that could be abandoned before a bus factor problem occurs, to estimate a percentage of files whose development could be easily continued by a new engineer without getting in touch with their colleagues, and to estimate the time scale in which the knowledge about an artifact changes by a factor of two.

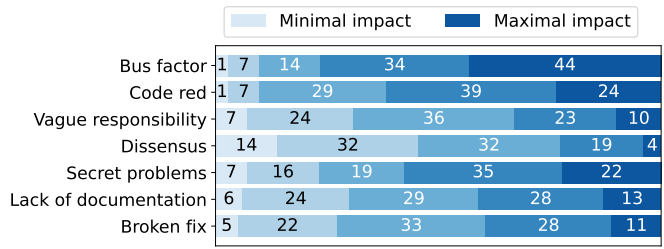

Figure 1: Perceived importance of collective development problems. All numbers are percentages

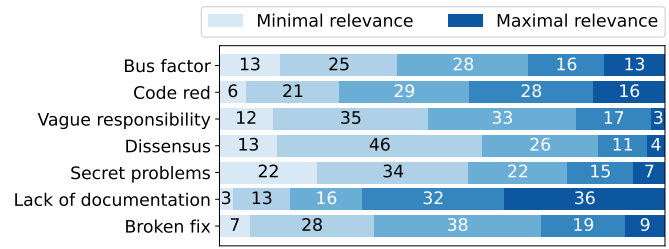

Figure 2: Perceived relevance of collective development problems. All numbers are percentages

The fourth part was related to the practical perception of the bus factor data. In it we have asked the engineers about the desired granularity for the bus factor data, the way they would like to get the results communicated to them, and possible project-related actions the engineers may take based on the bus factor data.

*Informed consent.* The survey included the informed consent form.

*Recruitment.* We recruited participants through Slack chats and JetBrains mailing lists. We have also posted the survey in the social media such as Twitter or LinkedIn and asked respondents to forward the survey to other engineers. External respondents participated in the $100 Amazon card raffle.

*Respondents.* In total, 269 respondents of those who have completed the survey have been working in IT in a code-related role.

*4.2.1 General demographics information.* 8% of the respondents have less than 3 years of experience in IT, and 75% of the respondents have more than 6 years of experience in IT. 74% of the respondents were involved in six or more projects during the course of their career. 77% of the respondents have experience of leading a group of people. Thus they were responsible for the project development and may have been interested in tracking project health. 81% of the respondents are developers at one of their roles.

*4.2.2 Bus factor relevance.* 51% of the survey participants know, what bus factor is, but only 19% of the respondents have ever worked on a project, where the bus factor was communicated to them. The bus factor of a project is perceived to be an important metric of project health: on a scale from one to five, 75% of the respondents rated its importance at 3 or higher, and 39% of the respondents have rated the importance at 4 or higher. This perception is bolstered by the experience: 63% of the respondents over the course of past year have worked at least on one project, for which they felt there was a high risk of the bus factor reaching zero.

Respondents perceive the bus factor as the most important collective development problem out of seven problems that were suggested in the survey (see 4.1.1 for the list of problems), see Figure 1. Code red collective development problem was perceived as the second most important problem. Bus factor and code red were ranked as the fourth and the second most often occurring problems (Figure 2), which, combined with their high impact as perceived by the respondents highlights the necessity to avoid them and bolsters the idea of bus factor being an important metric of the project health.

*4.2.3 Bus factor algorithm details.* The respondents ranked different options for contribution to a project according to their importance, the results are presented in Table 2.

| Contribution Mode | MRR |
| --- | --- |
| Commits | 0.560 |
| Code reviews | 0.403 |
| Issues (YouTrack / GitHub / Space issues) | 0.316 |
| Test cases | 0.296 |
| Project documentation | 0.268 |
| Code comments | 0.222 |
| Online meetings | 0.214 |
| Meetings in person | 0.203 |
| Open channels in Slack / Space / etc. | 0.149 |
| Closed team chats in Slack / Space / etc. | 0.146 |
| Direct messages in Slack / Space / etc. | 0.140 |
| Other, please specify in the comment below | 0.101 |
| Mailing lists | 0.092 |

Table 2: Importance of various modes for knowledge exchange and creation, as reported by the survey participants

The survey results support our choice of code reviews as one of the modes to be considered in the algorithm.

76% respondents believe that the knowledge decays with time, but there was little agreement for a typical time scale in which the knowledge about a particular file halves, see Figure 3(a). Reported median halving time was 4 months. Most respondents believe that less than 20% of the files can be abandoned before project gets into trouble, see Figure 3(b). There was also little agreement on what percentage of files could be easily developed even if every engineer who worked on them have left the project, see Figure 3(c). Different respondents want different level of granularity for the bus factor calculation: 39% of the respondents prefer project-level calculations, 16% prefer module-level calculations and 40% prefer team-level computations. All these disagreements may be explained by differences in the structure of various projects. Thus, a user of a bus factor tracking tool should be able to tune these parameters.

61% of the respondents would like to get information on names of key engineers and parts of critical (lower bus factor) modules of the project together with the bus factor number. IP1, IP3, IP6, IP11 argue that for the small projects everyone in the team usually knows the project status and the bus factor, and IP4, IP6, IP7, IP10

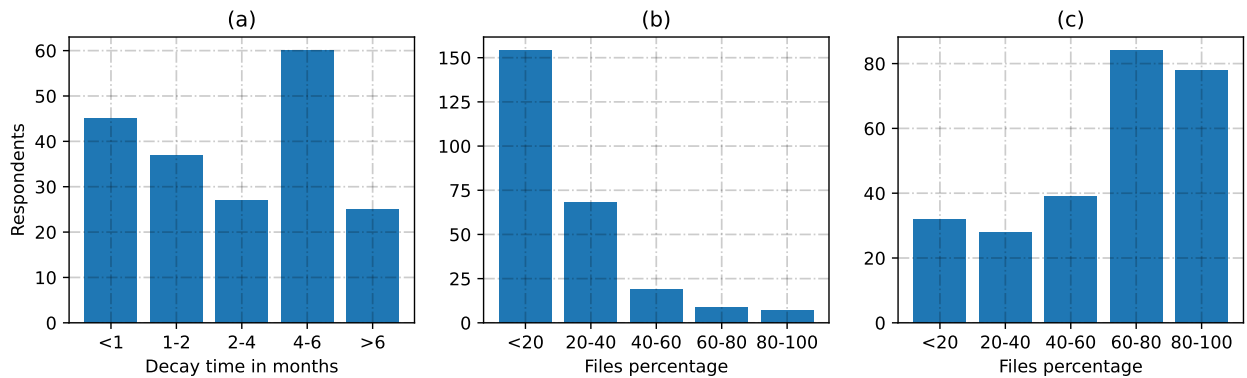

Figure 3: (a) - Decay time in which knowledge about file halves, (b) - Share of files that could be abandoned before project gets into trouble, (c) - Share of files that could be developed even if all engineers who worked on them have left

and some survey respondents reported that they already know the modules at risk. However, bus factor information may be useful to persuade management or team to pay more attention to a project or its certain parts, as mentioned by IP9 and IP10. IP1 and survey respondents has also mentioned they would like to be able to set manually the importance weights for the files.

*4.2.4 Practical use of the bus factor metric.* Both in the survey and in the interviews we have asked the respondents to weigh in on the ways bus factor data can be used. Respondents have suggested various ways they would address low bus factor, which include:

- Delving themselves in the parts at risk or finding other developers to work on them, trying to gain more knowledge about them (IP1, IP3, IP5, IP8, IP11, IP12).
- Attaching key engineers to the company by high salaries and good relationships (IP6, IP12).
- Refactoring and improving the code quality to make the code base more accessible.
- Designing the system so that the responsibilities are divided from the beginning (IP1, IP3, IP7, IP12).
- Organizing knowledge sharing on the parts at risk by writing documentation, organizing talks by key developers, doing code reviews, or including team into making decisions on the implicated parts of the project (IP2, IP4, IP10, IP11).
- Rearranging the workforce to ameliorate the risk by rotating people between the project parts, asking engineers who don't work on the parts at risk to work on them, or hiring new people to increase the knowledge redundancy (IP3, IP5, IP6, IP8, IP9). Hiring new people may be hard: IP9 reports once spending 3 years looking for a suitable engineer.

Mentioned as other information to report with bus factor were:

- Information about the documentation available for the low bus factor parts of the project
- Information on code complexity of the implicated parts.
- Information on criticality of the implicated parts (IP7, IP9).
- Some kind of competence graph to show which project parts are understood by various project members (IP2, IP10, IP12).
- Some kind of competence graph that would show which competences are required by the project (IP12).

- Highlighting modules at highest risk (IP1, IP4, IP8, IP9).
- Suggesting how to increase the bus factor (IP2, IP5).
- Information on risks that a particular developer is likely to leave the project (IP6).

*4.2.5 Pitfalls of using bus factor.* The bus factor metric and especially the key engineers part of it may not be received well by the team. Some of the concerns mentioned by the respondents include:

- Too much attention may be paid to the implicated modules.
- A key engineer may feel irreplaceable, which may negatively impact their relationship with their employer and teammates (IP1). One of the ways to ameliorate this is to make general bus factor data only available to the management, so that the engineers will know only the data about their direct participation (IP1). Another option is to support collective ego against the individual ego (IP11).
- Those not listed as key engineers may feel underestimated and uncomfortable about their role in the project.
- If the bus factor metric is taken at a face value, overestimation of the bus factor may result in false feeling of security.
- Engineers who work at the projects with low bus factor may be discouraged by the risk or even leave the project.

## 4.3 Bus factor tool validation survey

We have also conducted a survey to check whether the results of the tool we developed agree with the engineers' perception. This survey was the part of the bus factor general survey that was only available to the JetBrains employees.

*Survey protocol.* In the first question, the respondents provide a list of at most 3 projects they are working on. In the next question, the respondents estimate the bus factor and list the core developers for these projects. All survey questions were optional.

*Recruitment.* We recruited participants through Slack chats and the internal mailing lists of JetBrains. Respondents received no compensation for participating in the survey.

*Respondents.* In total, 14 respondents from 13 different projects have participated in our survey.

*Analysis.* For each of the 13 reported projects we compared the human estimates of the bus factor with the one provided by our tool and the baseline algorithm (Avelino et al. [5]). There was significance variance in reported bus factor for the large projects. To measure how close the predictions are to the ground truth provided by the survey respondents, we calculate mean absolute error (MAE). We got slightly better MAE than the baseline (5.46 and 5.80). We got answers on key engineers for eight projects. As the key engineers lists were somewhat different, we took the union of provided lists as the ground truth. The algorithms predicted the same key engineers with the same order in seven of them (primarily due to the lack of code review data), and the F1 score was 0.48 (P: 0.64, R: 0.39). In the remaining project, the algorithms ranked differently the first key developer suggested by the respondents. Our tool predicted them as a second most important engineer, while the baseline's predicted them to be third. We also analyzed the key developers predicted by the baseline algorithm in all 13 projects. We found that for two projects it predicted a key developer who is not active anymore (the last commit date was more than 1.5 years ago).

## 5 DISCUSSION

### 5.1 Contribution and novelty

The main contribution of this paper is the analysis of engineers' opinions and requests on the composition of the bus factor estimation algorithm and the corresponding tool. The analysis is further bolstered by a study of comparative significance and relevance of the collective development problems which shows that the bus factor and a related code red collective development problem are considered to be the most impactful collective development problems that also appear relatively often during project development.

Based on this analysis, we present the first multimodal bus factor assessment tool that estimates the bus factor and a list of key engineers using code reviews and meetings metadata in addition to the VCS data. In contrast, prior works on this subject [5, 10, 18, 19, 25] only account for VCS knowledge distribution mode. Our algorithm also accounts for the knowledge decay according to the forgetting curve, while the previous works either did not account for it [5, 10, 25] or considered a time window such that all the data created outside the window was discarded [13, 19].

Finally, we use exploratory interviews and surveys to derive a set of suggestions and requests for a bus factor assessment tool that may be used as a reference in creating other bus factor tools.

### 5.2 Insights

While respondents agree that the bus factor is a threat to be avoided, there is little agreement on how the tool should work. For example, while IP5 requests a tool that would suggest "let engineer X work on the subsystem Y", IP7 says that advice like that will be useless. Nevertheless, it is possible to extract several insights relevant for the creators of a bus factor assessment tool:

- The bus factor assessment tools in their current form (that address a bus factor of a particular repo) are probably most useful for medium-sized teams. The small teams, less than 6-8 people (as estimated by IP1, IP4) have little need in the tool since the scope of work done in a team is comprehensible. The large teams work on projects consisting of many modules, and the bus factor problem can happen when even a single critical module reaches bus factor zero.
- The team using the bus factor tool should be able to tune the parameters in the bus factor algorithm. Our survey results highlight that engineers have different opinions on the relative importance of various knowledge distribution modes or on the knowledge decay rate. While making the tool adjustable may result in users fitting tool to their biases, a one-size-fits-all solution cannot encompass all the diversity of different software development projects.
- As different engineers have different requests for the output of the tool, the tool report should be modular with the option to switch off a particular mode not desired by the user.
- The tool should never be perceived as an ultimate arbiter on the bus factor (or any other health metric) of a project. It should be used as an additional source of information on the project that allows condensing scattered information about the code ownership into a small description with an inevitable loss of context. This data can then be used to highlight the murky parts of the project, can be accompanied by the competence graph to highlight competencies lacking on the project, or can be presented as a supporting argument in a discussion on the project state.
- Even though the bus factor tool is not intended to be used to make any kind of personnel-related decisions, respondents were concerned about the concept and especially about the "key engineer" notion. The concerns were less pronounced during the interviews. We believe this difference means that the reasons and the use cases for the tool should be explained to the team. We also believe that the tool should be used on an opt-in basis per team, and the opt-in decision should involve the development team.

## 6 THREATS TO VALIDITY

There are several possible threats to the validity of our study.

Some of the survey respondents may not have enough experience to answer the questions. To mitigate this issue, we filtered out the answers by the respondents with no experience of working in IT.

It is possible that the survey results may not represent the opinion of the industry as a whole. To improve the generalizability of our results, we used external mailing lists and advertised our survey in social networks to gather more responses from the engineers who don't work at JetBrains. 243 out of 269 completed surveys were filled out by the engineers external to JetBrains.

Another threat is related to the researcher's bias when codifying the interviews and surveys. To mitigate the threat, the authors discussed the extracted categories.

Our validation study for the algorithm is based on a survey for the projects developed at a single enterprise company, which may not be a good representative of the industry practices in general.

The thresholds and parameters for the algorithm formula may have been chosen suboptimally. All the parameters were chosen before the survey was run, so there was no risk of unintentional coefficient overfitting. A better strategy would be to break the projects sample into train and test parts and fit the coefficients on the former; however, the sample size did not allow such an approach.

## 7 CONCLUSIONS AND FUTURE WORK

Bus factor is an important metric of project health that tracks one of the most impactful collective development problems. In this paper, we present a multimodal bus factor assessment tool that estimates the bus factor and the corresponding list of key project members from VCS history, code reviews and meetings metadata. We evaluate the quality of our tool on the dataset of projects developed at JetBrains to show an improvement of 0.34 MAE over the previous state-of-the-art tool of Avelino et al. [5].

We have analyzed 12 exploratory interviews and a survey of 269 engineers. We present a set of recommendations and requests for the design of a bus factor tool that is based on the results of these interviews and survey. Using the interviews and the survey, we derive a set of best practices to address the bus factor issue, and why the team may fail to adopt these practices. Based on these findings, we suggest use cases for the bus factor estimation tool.

In the future, we would like to extend the functionality of our tool to estimate the bus factor of separate modules in the project. We would like to add more projects to the evaluation set and split it into two parts to fit the coefficients of the algorithm on the train part and evaluate the algorithm on the test part of the dataset. We would also like to include additional modes of knowledge creation and distribution, such as documentation, issue trackers data, or test cases. We are also interested in doing an ablation study to gauge the relative importance of various modes considered by the tool and the significance of the forgetting curve factors. Finally, it would be interesting to augment the tool with additional metrics to provide a comprehensive picture of project health.

## 8 ACKNOWLEDGEMENTS

We thank Yanina Ledovaya, Elli Ponomareva, Maria Antropova, Anastassiya Sichkarenko, and Egor Akhmetzianov for their advice and help with the qualitative part of this study.